\documentclass[]{aa}  

\usepackage{graphicx}
\usepackage{txfonts}
\usepackage{soul}

\newcommand{\ltsima} {$\; \buildrel < \over \sim \;$}
\newcommand{\gtsima} {$\; \buildrel > \over \sim \;$}
\newcommand{\lta} {\lower.5ex\hbox{\ltsima}}
\newcommand{\gta} {\lower.5ex\hbox{\gtsima}}

\newcommand{\kms}{km\ s$^{-1}$}

\begin{document} 

\title{Outflow of hot and cold molecular gas from the obscured secondary nucleus of NGC\,3256: closing in on feedback physics}
\titlerunning{Outflow of hot and cold molecular gas from the obscured secondary nucleus of NGC\,3256}

\subtitle{}

   \author{B.\,H.\,C. Emonts
          \inst{1}\thanks{Marie Curie Fellow}
          \and J. Piqueras-L\'{o}pez
          \inst{1}
          \and L. Colina
          \inst{1}
          \and S. Arribas
          \inst{1}
          \and M. Villar-Mart\'{i}n
          \inst{1}
          \and M. Pereira-Santaella
          \inst{1,2}
          \and S. Garcia-Burillo
          \inst{3}
          \and A. Alonso-Herrero
          \inst{4}
          }

   \institute{Centro de Astrobiolog\'{i}a (INTA-CSIC), Ctra de Torrej\'{o}n a Ajalvir, km 4, 28850 Torrej\'{o}n de Ardoz, Madrid, Spain\\
              \email{bjornemonts@gmail.com}
         \and Istituto di Astrofisica e Planetologia Spaziali (INAF), Via Fosso del Cavaliere 100, I-00133 Roma, Italy
         \and Observatorio Astron\'{o}mico Nacional (OAN), Observatorio de Madrid, Alfonso XII, 3, 28014, Madrid, Spain
         \and Instituto de F\'{i}sica de Cantabria (CSIC-UC), 39005 Santander, Spain
             }

   \date{}

 
\abstract{The nuclei of merging galaxies are often deeply buried in dense layers of gas and dust. In these regions, gas outflows driven by starburst and active galactic nuclear activity are believed to play a crucial role in the evolution of these galaxies. However, to fully understand this process it is essential to resolve the morphology and kinematics of such outflows. Using near-infrared integral-field spectroscopy obtained with SINFONI on the Very Large Telescope, we detect a kpc-scale structure of high-velocity molecular hydrogen (H$_{2}$) gas associated with the deeply buried secondary nucleus of the infrared-luminous merger-galaxy NGC\,3256. We show that this structure is most likely the hot component of a molecular outflow, which is detected also in the cold molecular gas by Sakamoto et al. This outflow, with a total molecular gas mass of M$_{\rm H_2}$$\sim$2\,$\times$\,10$^{7}$ M$_{\odot}$, is among the first to be spatially resolved in both the hot molecular H$_{2}$ gas with VLT/SINFONI and the cold molecular CO emitting gas with ALMA. The hot and cold components share a similar morphology and kinematics, with a hot-to-cold molecular gas mass ratio of $\sim$\,$6$\,$\times$\,10$^{-5}$. The high ($\sim$100 pc) resolution at which we map the geometry and velocity structure of the hot outflow reveals a biconical morphology with opening angle $\sim$40$^{\circ}$ and gas spread across a FWZI\,$\sim$\,1200 \kms. Because this collimated outflow is oriented close to the plane of the sky, the molecular gas may reach maximum intrinsic outflow velocities of $\sim 1800$ \kms, with an average mass outflow rate of at least \.{M}$_{\rm outfl} \sim 20$ M$_{\odot}$\,yr$^{-1}$. By modeling the line-ratios of various near-infrared H$_{2}$ transitions, we show that the H$_{2}$-emitting gas in the outflow is heated through shocks or X-rays to a temperature of T\,$\sim$\,$1900\pm300$\,K. The energy needed to drive the collimated outflow is most likely provided by a hidden Compton-thick AGN or by the nuclear starburst. We show that the global kinematics of the molecular outflow that we detect in NGC\,3256 mimic those of CO-outflows that have been observed at much lower spatial resolution in starburst- and active galaxies. 
}

   \keywords{galaxies: NGC 3256 -- galaxies: starburst -- galaxies: active -- galaxies: nuclei -- ISM: jets,\,outflows -- ISM: dust,\,extinction}

   \maketitle

%

\section{Introduction}
\label{sec:intro}

Luminous and ultra-luminous infrared galaxies, also called LIRGs ($L_{\rm IR}$ > 10$^{11}$ $L_{\odot}$) and ULIRGs ($L_{\rm IR}$ > 10$^{12}$ $L_{\odot}$), are galaxies with massive dust-enshrouded star formation and often a deeply buried active galactic nucleus \citep[AGN;][]{san96}. Because (U)LIRGs can be found at relatively low $z$, they are excellent laboratories for studying physical processes that are crucial in the evolution of massive galaxies, from galaxy merging to starburst/AGN-induced feedback. These processes, in particular when they occur in the nuclei, are often hidden behind thick layers of gas and dust, which absorb most of the light before re-radiating it at IR wavelengths.

Integral-field spectroscopy in the near infrared (IR) provides an excellent tool to study the physical processes in the central regions of (U)LIRGs. At near-IR wavelengths the dust extinction is much lower than in the optical. In addition, in the near-IR a variety of emission lines from different gas phases, as well as stellar absorption lines, can often be observed simultaneously (from coronal [Si\,VI] and ionized Br$\gamma$ or Pa$\alpha$ to partially ionized [Fe\,II] and molecular H$_{2}$ emission). Particularly interesting is that a range of near-IR transitions of H$_{2}$ can be targeted to study the physical properties of the molecular gas. Another intrinsic advantage of near-IR integral-field spectroscopy is that, apart from being able to image the gas at high spatial resolution, different gas structures can also be distinguished kinematically. This can be done in detail by decomposing complex line profiles into multiple kinematic components across the field of view of the spectrograph.

The strength of a detailed kinematic analysis of near-IR integral-field spectra was recently shown by \citet{rup13}, who revealed the base of a deeply buried H$_{2}$ outflow in the nearby quasi-stellar object QSO F08572+3915. Other observational techniques have also revealed evidence that links nuclear AGN and starburst activity with heating and outflow of neutral and molecular gas in low-$z$ galaxies \citep{oos00,leo07,fer10,ala11,rup11,aal12,gui12,das12,com13,mor05,mor13sci,mor13,mah13,das14,gar14,caz14,tad14}. Moreover, recent surveys with instruments like Herschel and the Plateau de Bure Interferometer revealed evidence of massive outflows of molecular gas in ULIRGs \citep[e.g.,][]{chu11,stu11,spo13,vei13,cic14}, which complement extensively studied ionized gas outflows in these systems \citep[e.g.,][]{hec90,wes12,bel13,rod13,arr14}. Thus, if feedback onto the molecular gas is common in merger systems with dust-obscured starburst/AGN cores, we expect to find evidence of this through detailed 3D kinematic studies of the near-IR H$_{2}$ lines in other nearby IR-luminous galaxies.

In this paper, we perform a kinematic and morphological analysis of the near-IR H$_{2}$ emission from molecular gas associated with the deeply buried secondary nucleus of NGC\,3256, which is the most luminous LIRG within $z < 0.01$ \citep[$L_{\rm ir} \sim 5 \times 10^{11} L_{\odot}$;][]{san03}. In \citet{piq12} we previously revealed that NGC\,3256 contains a large-scale rotating H$_{2}$ disk, which we imaged by tracing the H$_{2}$\,1-0\,S(1) $\lambda$=2.12$\mu$m line with near-IR integral-field data from the {\sl Spectrograph for Integral Field Observations in the Near Infrared at the Very Large Telescope} \citep[VLT/SINFONI;][]{eis03,bon04}. However, the standard single-Gaussian fitting procedure that we applied in \citet{piq12} revealed residuals that suggested the presence of more complex kinematics across various regions (see Sect.\,\ref{sec:data}). We further explore the complex kinematics of the H$_{2}$ gas in NGC\,3256 in this paper.

\subsection{NGC\,3256}
\label{sec:intro3256}

NGC\,3256 is a gas-rich merger with prominent tidal tails, galactic winds and ionized gas outflows \citep[e.g.,][]{sca96,mor99,hec00,lip00,lip04,eng10,mon10,ric11,lei13,bel13,arr14}. It contains two nuclei, separated by $\sim$1\,kpc. The secondary, or southern, nucleus is heavily obscured, as revealed with IR and radio observations \citep[][]{nor95,kot96,alo06a,lir08,san08}. Various authors have claimed that this secondary nucleus may host a heavily obscured AGN \citep[e.g.,][]{kot96,nef03}. So far, X-ray studies provided inconclusive evidence for this \citep{lir02,jen04,per11}, while the near- and mid-IR properties of both nuclei are ascribed mostly to star-formation activity \citep{alo06b,lir08,per10,alo12}.

Molecular gas was detected through observations of H$_{2}$ and various tracers of cold molecular gas \citep[e.g.,][]{moo94,doy94,sar89,mir90,aal91,cas92,baa08}. \citet{sak06} used CO(2-1) observations taken with the Submillimeter Array (SMA) to map a large ($r$\,>\,3\,kpc) disk of cold molecular gas rotating about the mid-point between the two nuclei. Part of this disk was also mapped in H$_{2}$ \citep{piq12} and H$\alpha$ \citep{lip00,zau11,bel13}. \citet{sak06} also found a high-velocity component of molecular gas. In a concurrent paper, \citet{sak14} present new ALMA data that reveal that the high-velocity molecular gas is associated with two outflows, namely a starburst-driven superwind from the primary nucleus and a much more collimated kpc-scale bipolar outflow from the secondary nucleus. Our VLT/SINONI H$_{2}$ data reveal independent evidence of this collimated bipolar outflow and will shed a new light on its nature.

Throughout this paper, we assume $z=0.009354$ (which is the redshift of the Br$\gamma$ peak emission at the location of the secondary nucleus) and D\,=\,44.6 Mpc (1$''$ = 216 pc), as per \citet{piq12}.


\section{Data}
\label{sec:data}

\begin{figure*}
   \centering
   \includegraphics[width=0.86\hsize]{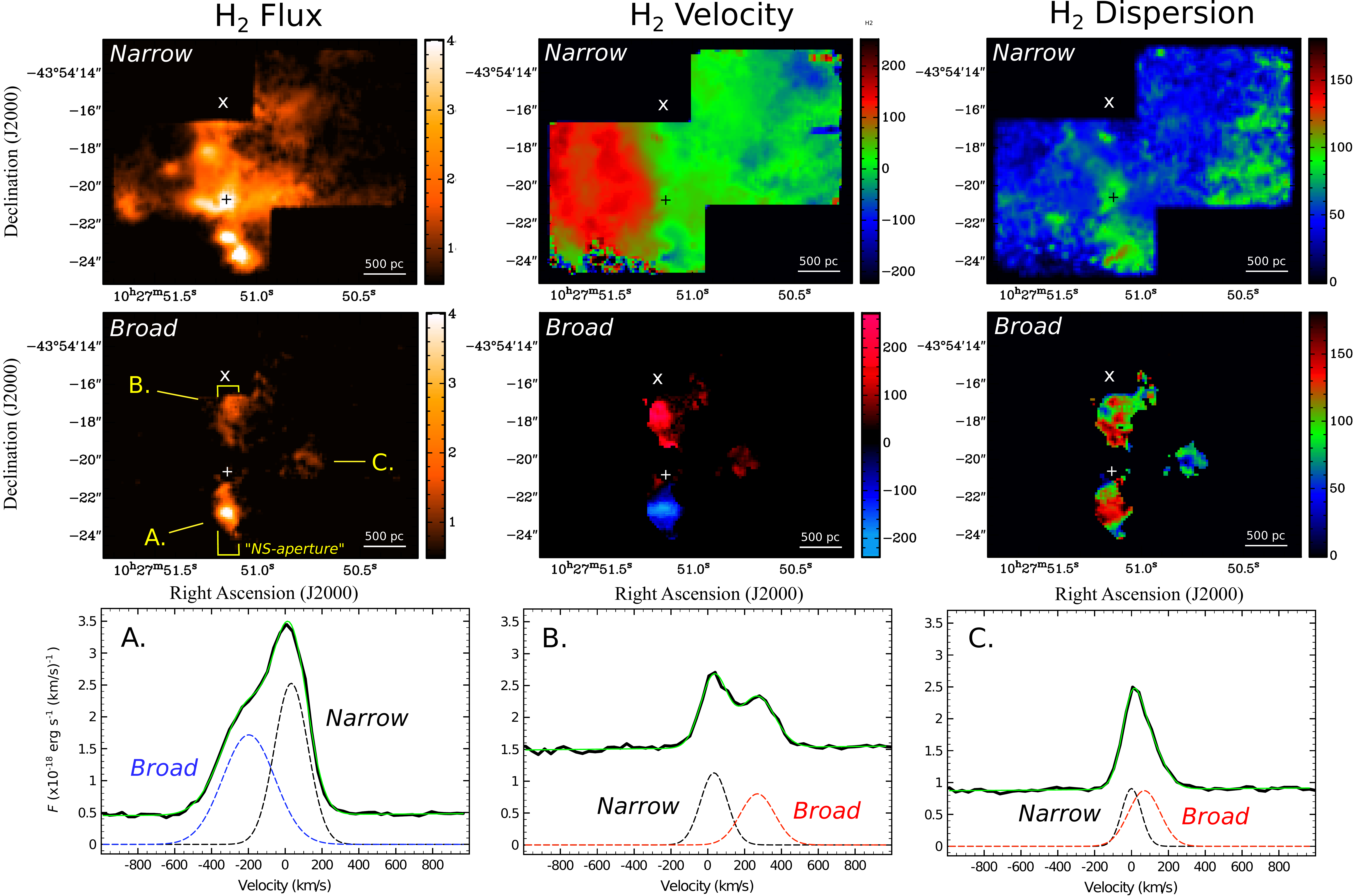}
      \caption{SINFONI view of the H$_{2}$ emission in NGC\,3256, mostly covering the region surrounding the secondary (southern) nucleus. The primary and secondary nucleus are marked with an $\times$ and +, respectively, and are shown in more detail in Fig.\,\ref{fig:NICMOS}. {\sl Top + middle:} Maps of the H$_{2}$ flux ($F$), velocity of the emission-line peak (v) and velocity dispersion ($\sigma$\,=\,FWHM/2.35, with FWHM the full width at half the maximum intensity) of both the narrow- {\sl (top)} and broad-component gas {\sl (middle)}. Velocities are with respect to $z=0.009354$ as per \citet{piq12}, which is the assumed systemic velocity of the secondary core (Sect.\,\ref{sec:intro3256}). Units are in $\times 10^{-17}$ erg\,s$^{-1}$\,cm$^{-2}$ ($F$) and \kms\ (v and $\sigma$). {\sl Bottom:} H$_{2}$ emission-line profiles, plus continuum-subtracted 2-component Gaussian fit, in regions A, B and C. The spectra were extracted from a circular aperture of 13 spaxels of $0.125'' \times 0.125''$ each (i.e., roughly the size of the seeing disk), centered on the broad-component H$_{2}$ features.}
         \label{fig:data}
   \end{figure*}

We use near-IR data obtained with VLT/SINFONI previously presented in \citet{piq12}, with pixel-size 0.125$''$, seeing $\sim$0.6'', spectral resolution 6.0$\pm$0.6\,\AA\ and dispersion 2.45\,\AA/pixel. These data originally revealed a large-scale H$_{2}$ disk (Sect.\,\ref{sec:intro}), which is also shown in our Fig.\,\ref{fig:data}. From visual inspection, the single Gaussian fit previously applied to the H$_{2}$\,1-0\,S(1) emission line by \citet{piq12} did not produce satisfactory results for mapping the full kinematic structure of the hot molecular gas. This was most notable in the regions $\sim$$2.5$ arcsec south and $\sim$3.5 arcsec north of the secondary nucleus in \citet{piq12}, where the measured velocity from the Gaussian fit to the H$_{2 }$\,1-0S\,(1) line did not reflect the otherwise regular rotation pattern of the large-scale H$_{2}$ disk. For the current work, we modified the data analysis routine used by \citet{piq12} to fit two Gaussian components to the H$_{2}$ line profile there where required. This routine was compiled for the Interactive Data Language (IDL) and uses the MPFIT package for the $\chi^{2}$ minimization \citep{mar09}. No constrains were placed on the fitting parameters and all fits were visually inspected to ensure their validity. A flux ($F$), velocity (v) and velocity dispersion ($\sigma$) map (corrected for instrumental broadening) were created for each component, and boxcar-smoothed by 3 pixels in both spatial directions. 

We also used archival ALMA cycle-0 data to compare the morphology and kinematics of certain H$_{2}$ features with those observed in CO(3-2) (program 2011.0.00525; PI K. Sakamoto, see also \citealt{sak13a,sak13b}). These data have a spatial resolution of $1.2'' \times 0.7''$ (PA = 23$^{\circ}$) and channel width of 10 \kms, with the standard ALMA achival pipeline-reduction applied. In a concurrent paper, \citet{sak14} present these CO data (and other molecular gas tracers) using additional calibration and analysis techniques, hence we refer to their work for further details on the CO properties of NGC\,3256.

Images of the Hubble Space Telescope {\sl Near Infrared Camera and Multi-Object Spectrometer 2} (HST/NICMOS2) filters F160W, F212N and F237M (prog.\,ID 7251, PI: J.\,Black; see also \citealt{ros07}) and the {\sl Wide Field and Planetary Camera 2} (HST/WFPC2) filter F814W (prog.\,ID 5369, PI: S.\,Zepf; see also \citealt{zep99}, \citealt{lai03}) were extracted through the Hubble Legacy Archive. Our SINFONI data were overlaid onto these HST images by visually matching the coordinates to those of the secondary nucleus in the HST/NICMOS2 data.

   \begin{figure}
   \centering
   \includegraphics[width=0.8\hsize]{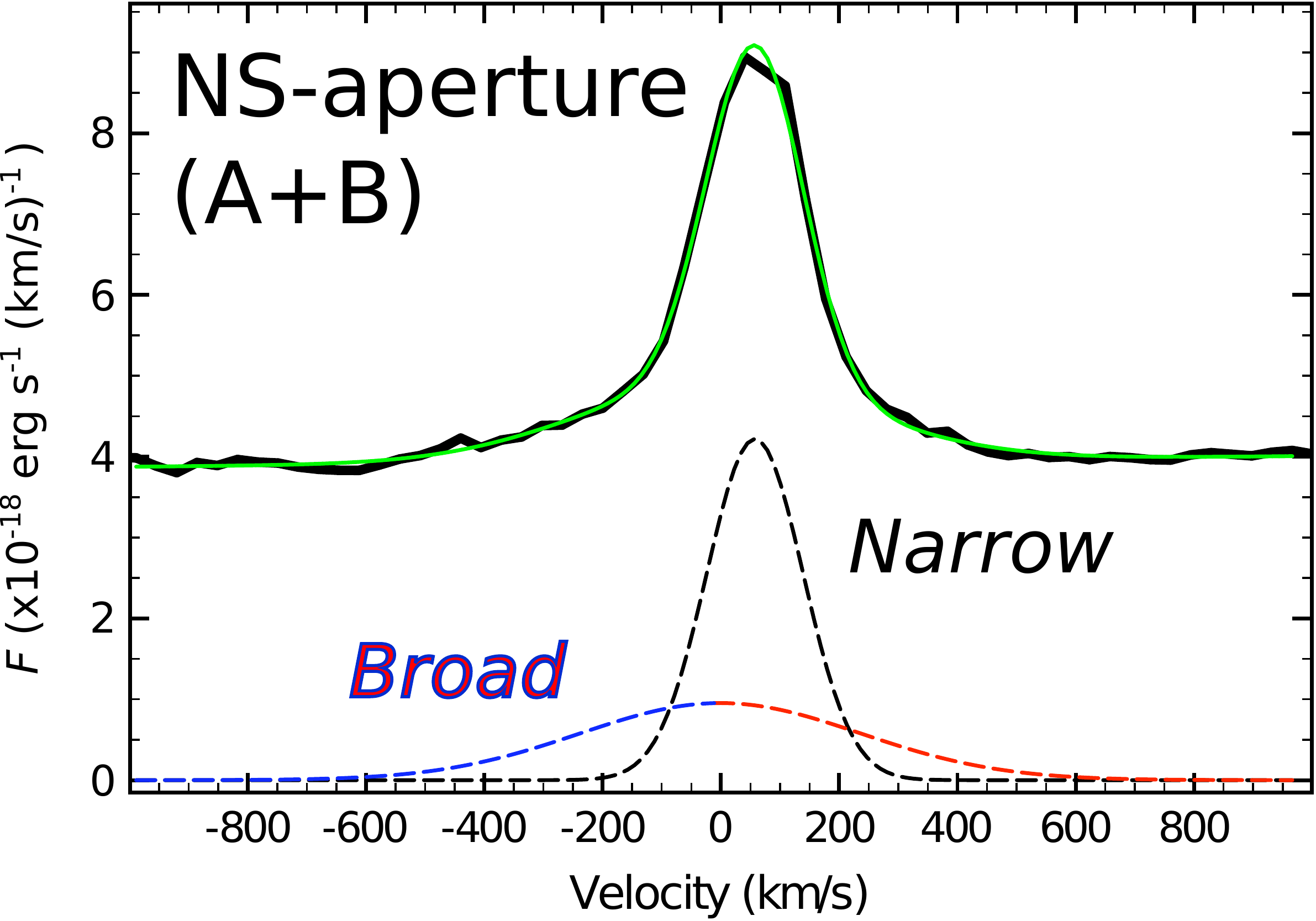}
      \caption{Integrated H$_{2}$ emission-line profile across the combined 1.1$''$ $\times$ 8.8$''$ north-south aperture that covers the full broad-component A+B region (as visualized in the {\sl middle-left} panel of Fig.\,\ref{fig:data}). The 0-velocity is defined as the redshift of the Br$\gamma$ peak-emission at the location of the secondary nucleus. The $\sim$60 \kms\ redshift of the narrow component is likely due to the fact that the kinematic center of the large-scale molecular disk is $\sim$500 pc north of the secondary nucleus.}
         \label{fig:datatotal}
   \end{figure}

\section{Results}

\subsection{Hot molecular structure}
\label{sec:results}

The results of mapping the H$_{2}$\,1-0\,S(1) line with two Gaussian components are shown in Fig.\,\ref{fig:data}. A ``narrow'' component is visible across the field of view (F.o.V.), tracing the large-scale rotating molecular gas disk previously discussed by \citet{piq12}. A second, ``broad'' component is detected in three regions (A, B and C in Fig.\,\ref{fig:data}). Properties of the broad-component emission are given in Table \ref{tab:results}.

The most prominent broad-component features are the bright, blue-shifted structure A (projected distance $\sim 440$\,pc south of the secondary nucleus) and the redshifted structure B ($\sim 800$\,pc north of the secondary nucleus). The integrated intensity is $I$\,=\,4.2 and 3.8 $\times 10^{-15}$\,erg\,s$^{-1}$\,cm$^{-2}$ across regions A and B, respectively. This H$_{2}$ gas has velocities that appear to be decoupled from those of the main narrow-component disk and shows a significantly higher velocity dispersion than the disk emission. When this broad component is integrated across the full extent of regions A+B, it appears to be centered both spatially and kinematically around the secondary nucleus. The corresponding integrated broad-component profile shown in Fig.\,\ref{fig:datatotal} covers a total full-width-at-zero-intensity velocity range of FWZI$_{\rm tot}$ $\approx$ 1200 \kms\ (or FWHM$_{\rm tot}$ $\sim$ 560 \kms) and is symmetric with respect to our assumed systemic velocity of the secondary nucleus. We note that our SINFONI F.o.V. does not cover the primary nucleus (north of region B), while the line-fitting south of region A becomes too uncertain (Fig.\,\ref{fig:data} {\sl top-middle}), hence the full extent of the broad-component H$_{2}$ structure may not be revealed.

   \begin{figure*}
   \centering
   \includegraphics[width=\hsize]{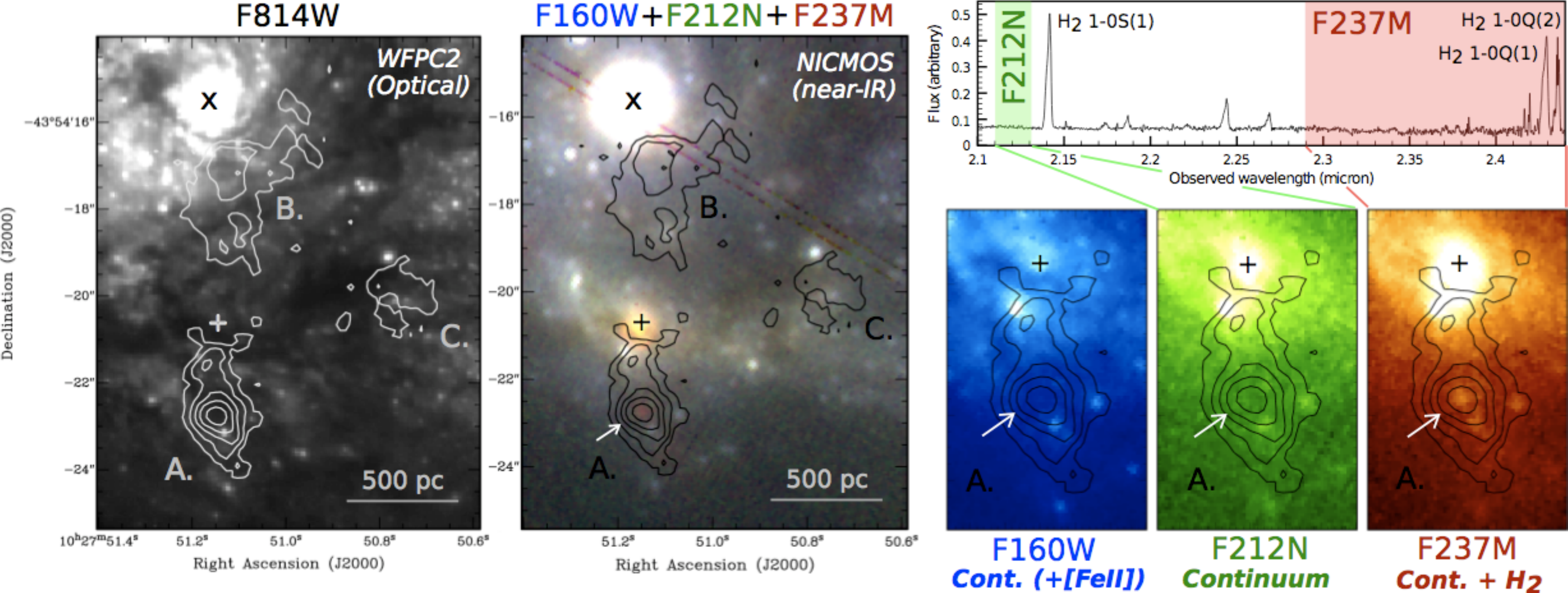}
      \caption{{\sl Left:} Contours of the broad-component H$_{2}$ emission from Fig.\,\ref{fig:data} (levels: 0.7, 1.3, 2.0, 2.8, 4.0 $\times 10^{-17}$ erg\,s$^{-1}$\,cm$^{-2}$) overlaid onto an optical image taken with the HST/WFPC2 F814W$_{\rm 724-878\,nm}$ wide-band filter. {\sl Middle:} Contours of the broad-component H$_{2}$ emission overlaid onto a 3-color near-IR image of the same region taken with the HST/NICMOS2 (filters: blue\,=\,F160W$_{1.4-1.8\,\mu m}$, green\,=\,F212N$_{\rm 2.111-2.132\,\mu m}$, red\,=\,F237M$_{2.29-2.44\,\mu m}$). {\sl Top right:} VLT/SINFONI spectrum obtained against the region of peak H$_{2}$ emission in region A (aperture 13 pixels). Visualized are the regions of the HST/NICMOS F212N narrow-band (2.111\,-\,2.132\,$\mu$m) and F237M medium-band filter (2.29\,-\,2.44\,$\mu$m). The F237M filter includes the prominent H$_{2}$\,1-0\,Q(1) and Q(2) lines. The F160W wide-band filter (1.4\,-\,1.8\,$\mu$m) covers the VLT/SINFONI H-band spectrum of this object, with [Fe\,II] as the most prominent emission-line in this spectral region \citep[see][]{piq12}. {\sl Right bottom:} Region A as observed in the three different NICMOS filters. The arrows point to the region with enhanced H$_{2}$ emission in both the VLA/SINFONI and HST/NICMOS\,F237M data.}
         \label{fig:NICMOS}
   \end{figure*}

Figure \ref{fig:NICMOS} shows a contour map of the broad-component H$_{2}$ emission overlaid onto various optical and near-IR HST images. The 3-color composite HST/NICMOS F160W+H212N+F237M image in Fig.\,\ref{fig:NICMOS} {\sl (middle)} shows a distinctive red feature at the location of the bright H$_{2}$ emission in region A. As shown on the right of Fig.\,\ref{fig:NICMOS}, this red feature (marked with an arrow) only appears in the NICMOS F237M filter. The F237M filter (2.29 - 2.44 $\mu$m) includes the H$_{2}$\,1-0Q(1), 1-0Q(2) and 1-0Q(3) emission lines at the redshift of NGC\,3256, of which the prominent Q(1) and Q(2) lines fall within the wavelength range of our SINFONI data.\footnote{Because of residual sky-line effects at the very edge of the SINOFONI wavelength coverage, we could not do an accurate spectral analysis of the H$_{2}$\,1-0Q(1) and H$_{2}$\,1-0Q(2) lines.} The same red feature does not appears in the F160W and F212N filters that mostly trace the near-IR continuum. We thus believe that the red feature south of the secondary nucleus in the NICMOS image consists of molecular line emission, which we also see as enhanced H$_{2}$ emission in both the narrow and broad component in region A (Fig.\,\ref{fig:data}). No obvious features are visible in any of the NICMOS filters in region B. We also find no obvious stellar or dusty counterpart that follows the broad-component H$_{2}$ emission across regions A and B in optical wide-band HST/WFPC2 imaging (Fig.\,\ref{fig:NICMOS}\,-\,{\sl left}), although a dark line does appear to cross the region of peak H$_{2}$ brightness (at least in projection). In addition, the Br$\gamma$-line of ionized gas in our SINFONI data is very weak in region A and does not have a prominent broad component in region B. It should be noted that the weak Br$\gamma$ does appear to have an increased velocity dispersion in region A \citep{piq12}, while a weak broad-component feature in H$\alpha$ has been reported south of region A by \citet{bel13}. Nevertheless, the broad component features in regions A and B appear to {\sl predominantly consist of molecular gas}.

A faint reservoir of redshifted H$_{2}$ is also seen in region C. This gas has $\sigma$ similar to that of the narrow-component disk, while the velocity shift between the narrow and broad component is much smaller than for regions A and B. This makes it difficult to perform an in-depth analysis of the H$_{2}$ emission in region C (as will become clear also from Fig.\,\ref{fig:app}). Moreover, the broad-component emission in region C is found along a prominent tidal arm in the NICMOS F237M data (Fig.\,\ref{fig:NICMOS}) and \citet{piq14} shows that -- unlike regions A and B -- region C coincides with a starforming region with bright Br$\gamma$ emission. In this paper we will therefore only present the basic H$_{2}$ results for region C, while a more detailed analysis of the near-IR emission-line properties in this and other starforming regions in nearby (U)LIRGs is left for a future paper (Piqueras et al in prep).

\begin{table}
\caption{Physical parameters of the broad-component H$_{2}$\,1-0S(1) features. $I$ is the integrated intensity of the H$_{2}$ emission. $\Delta$v$_{\rm peak}$ is the maximum velocity of the peak of the broad component with respect to $z=0.009354$ across each region. $\sigma$$_{\rm max}$ is the maximum velocity dispersion of the gas, corresponding to $\sigma$\,$\equiv$\,FWHM/(2$\sqrt{2\,{\rm ln}(2)}$, with FWHM the full width at half the maximum intensity of the broad component. 
The maximum values have been obtained from the images shown in Fig.\,\ref{fig:data}.}
\label{tab:results}      
\centering          
\begin{tabular}{lccc}     
 & Region A. & Region B. & Region C. \\
\hline                    
$I$ ($\times 10^{-15}$\,erg\,s$^{-1}$\,cm$^{-2}$) & 4.2 & 3.8 & 1.1 \\
$\Delta$v$_{\rm peak}$ (km\,s$^{-1}$) & -240 & +270 & +110 \\
$\sigma$$_{\rm max}$ (km\,s$^{-1}$) & 190 & 180 & 100 \\
FWHM$_{\rm max}$ (km\,s$^{-1}$) & 447 & 424 & 235 \\
\end{tabular}
\end{table}

   \begin{figure*}
   \centering
   \includegraphics[width=0.9\hsize]{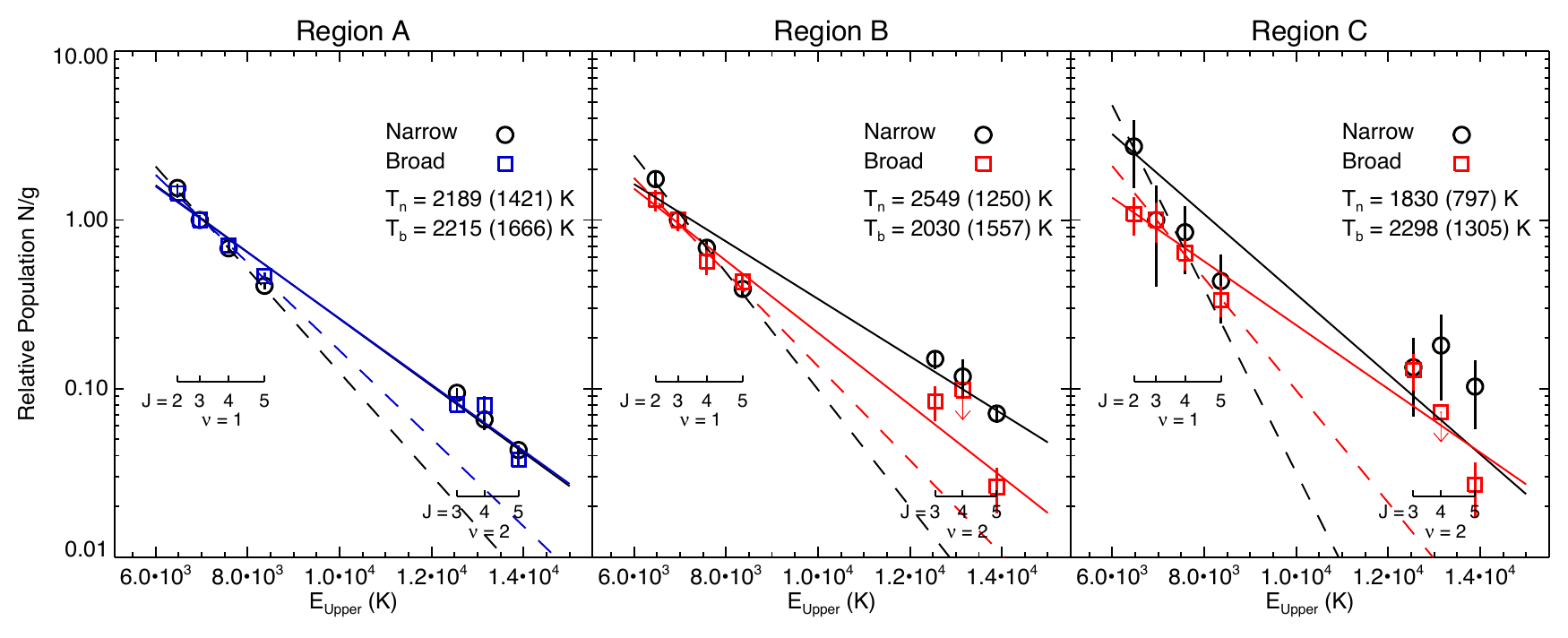}
      \caption{Modeling the relative population levels of the H$_{2}$ transitions using single excitation-temperature LTE models \citep[see][]{dav03}. The relative population levels have been derived from the intensities of the H$_{2}$ transitions listed in Table \ref{tab:app}. The circles and squares represent the narrow and broad component, respectively. The solid lines show the single-temperature fits using both the $\nu=1$ and $\nu=2$ transitions (i.e., assuming fully thermalized LTE gas conditions). The dashed lines show the single-temperature fits to only the $\nu=1$ transitions (i.e., taking into consideration that the $\nu=2$ transitions could be over-populated compared to fully thermalized LTE conditions). The best-fit temperatures for the narrow (n) and broad (b) component in the various regions are also show (with in brackets the temperature derive from the fit to only the $\nu=1$ transitions).}
      \label{fig:tempfits}
   \end{figure*}

   \begin{figure}
   \centering
   \includegraphics[width=\hsize]{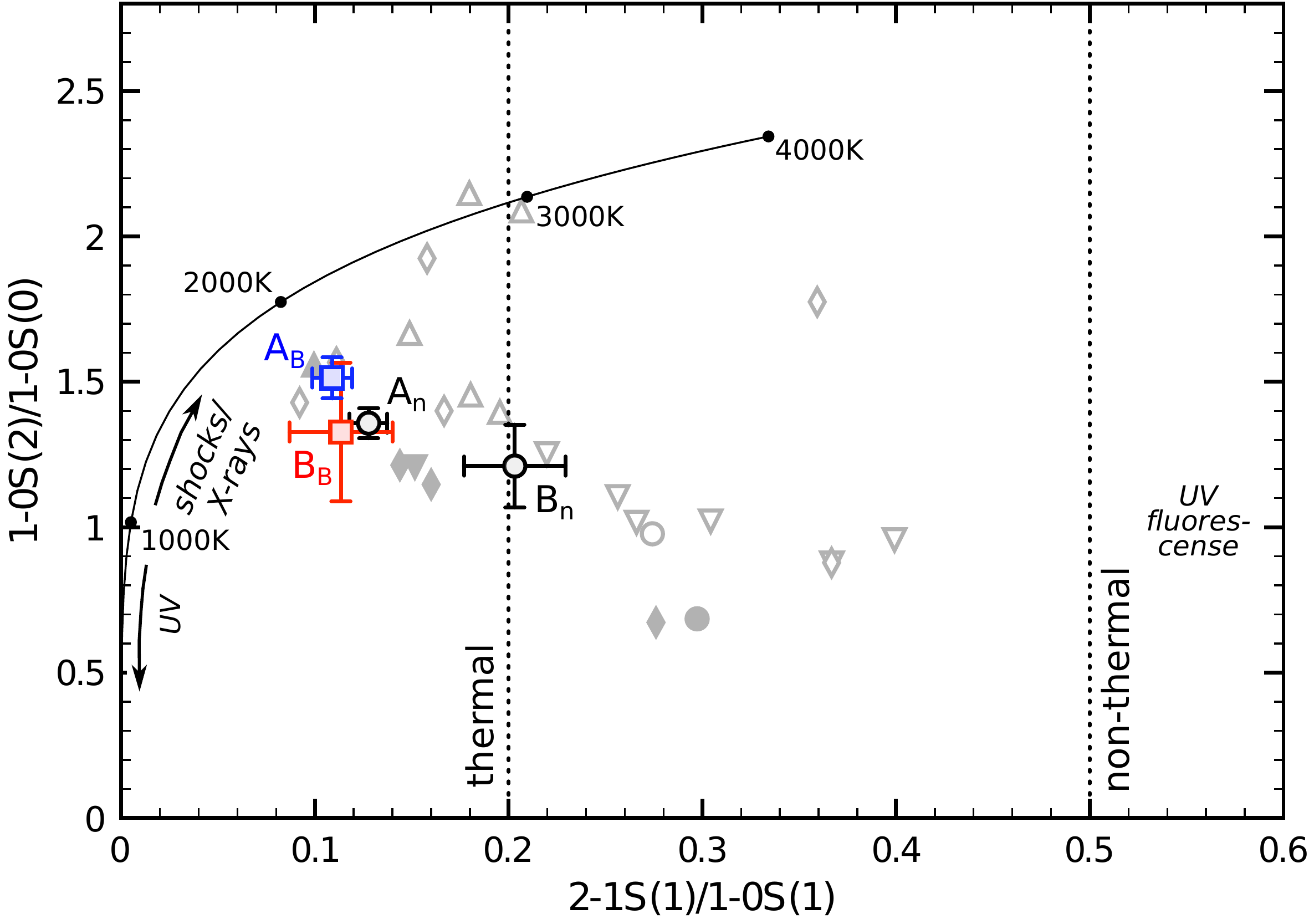}
      \caption{H$_{2}$ line ratio diagram for the blueshifted broad component in region A (blue square), the redshifted broad component in region B (red square) and the narrow components in regions A and B (black circles). The solid black line shows the expected line ratios for uniform density gas that is in Local Thermal Equilibrium (LTE), with excitation temperatures indicated. The dotted lines provide an indication for the regions where the 2-1\,S(1)/1-0\,S(1) line ratio is caused by thermal ($\lesssim$\,0.2) and non-thermal ($\gtrsim$\,0.5) conditions, based values from models that are summarized in \citet{mou94}. The regions where these models predict UV photons \citep{ste89}, shocks/X-rays \citep{bra89,lep83,dra90,dor12} and UV-fluorescence \citep{bla87} to dominate the gas excitation are indicated in more detail. Open gray symbols represent values from individual regions studied within other nearby galaxies (starforming regions, rings, arcs, clouds, etc.), with closed gray symbols the nuclei of these systems. Diamonds are from \citet{maz13}, downward pointing triangles from \citet{fal14}, upwards pointing triangles from \citet{bed09} and circles from \citet{piq12M83}. See also \citet{ram09} and \citet{rif13} for studies performed on integrated near-IR long-slit spectra of various types of galaxies.}
      \label{fig:H2ratios}
   \end{figure}

Apart from the H$_{2}$ 1-0\,S(1) 2.12$\mu$m line, we see a broad component also in other H$_{2}$ lines, namely H$_{2}$ 1-0\,S(0) 2.22$\mu$m, 1-0\,S(2) 2.03$\mu$m, 1-0\,S(3) 1.96$\mu$m, 2-1\,S(1) 2.25$\mu$m, 2-1\,S(2) 2.15$\mu$m and 2-1\,S(3) 2.07$\mu$m. For all H$_{2}$ transitions we performed a kinematic analysis similar to that of the 1-0\,S(1) line. Details of this procedure, as well as the line profiles and resulting line fluxes, are provided in Appendix \ref{app:H2lines}.


\subsubsection{Temperature and excitation mechanism}
\label{sec:temperature}

The relative intensities of the various H$_{2}$ transitions can be used to discriminate between the various mechanisms that excite the H$_{2}$ gas, and to estimate the gas temperature. Models show that H$_{2}$ can be excited either through thermal processes, such as shocks or UV/X-ray radiation, or non-thermal processes like UV-fluorescence \citep[e.g.,][]{lep83,hol89,ste89,bra89,dra90,mal96,dor12,bla87}. In case of thermal processes, the H$_{2}$ levels are excited through inelastic collisions, resulting in gas that is approximately in Local Thermal Equilibrium (LTE). For gas in LTE the various H$_{2}$ line ratios can be fitted by a single excitation-temperature model \citep[see, e.g.,][]{dav03}. In the case of non-thermal excitation, absorption of UV photons with $\lambda > 912$\AA\ is followed by a fluorescent cascade of electrons through the vibrational and rotational levels of the ground state. In this case, even though in most environments the lowest vibrational levels of H$_{2}$ ($\nu = 1$) are still well thermalized, the higher level transitions are significantly stronger than expected based on single-temperature LTE models and more complex models are required to derive the gas temperature \citep{dav03}. 

To estimate the excitation temperature of the H$_{2}$-emitting gas, we fit LTE models to the relative population levels across the range of H$_{2}$ transitions for the various components (see Appendix \ref{app:H2lines}). As can be seen in Fig.\,\ref{fig:tempfits}, if a single-temperature model is fitted to both the $\nu = 1$ and $\nu = 2$ transitions, we derive an estimated temperature of T\,$\sim$\,$2000-2200$\,K. A fit to only the $\nu = 1$ transitions renders a temperature of T\,$\sim$\,$1600$\,K, in which case the $\nu = 2$ transitions are slightly over-populated. This indicates that the excitation conditions of the broad-component H$_{2}$ gas may be somewhat more complex than single-temperature LTE conditions, with non-thermal processes adding to the gas excitation, multiple gas temperature components, or a low-density, sub-thermally excited gas phase mixed with denser LTE-gas. However, our coverage of only the $\nu = 1$ and $\nu = 2$ transitions, of which the $\nu = 2$ lines are relatively weak (Fig.\,\ref{fig:app}), does not allow us to constrain the model-fitting in great detail. Moreover, as we will see in Sect.\,\ref{sec:ALMA}, the broad-component emission is also seen in the cold molecular CO(3-2) and CO(1-0) gas with an estimated ratio that suggests that also the cold gas may be thermalized. We therefore here adopt the temperature of T\,$\sim$\,$1900\pm300$\,K for the broad-component emission-line gas in regions A and B.

The bright emission from the narrow-component H$_{2}$ disk in regions A and B shows roughly similar temperature-estimates when fitting LTE models. However, for this narrow component, the over-population of the $\nu = 2$ transitions is more prominent than for the broad-component emission, in particular in region B. This suggests that the temperatures of the narrow-component H$_{2}$ gas are more uncertain, and that the excitation conditions in parts of the large-scale disk are likely more complex, than those of the broad-component emission. However, a detailed discussion of this is beyond the scope of this paper. For both the narrow- and broad-component emission in region C the errors in the fluxes for the various transitions are too large to derive any meaningful conclusions.

We use the H$_{2}$ line ratios to further investigate the excitation-mechanism that dominates the heating of the H$_{2}$ gas. Following \citet[][see also \citealt{tan89}]{mou94}, we plot the ratio of the ortho-transitions 2-1\,S(1)/1-0\,S(1) against the ratio of the para-transitions 1-0\,S(2)/1-0\,S(0), which allows us to estimate whether thermal or non-thermal processes dominate the gas excitation. The results shown in Fig.\,\ref{fig:H2ratios} indicate that the gas-heating of the various components in regions A and B is dominated by thermal processes.\footnote{The line ratios for the narrow and broad component in region C appear consistent with this scenario, but the large errors (Fig.\,\ref{fig:tempfits}) do not allow us to derive firm conclusions, hence region C has been omitted from Fig.\,\ref{fig:H2ratios}.} More specifically, the 1-0\,S(2)/1-0\,S(0) ratio of the various components (plotted on the y-axis of Fig.\,\ref{fig:H2ratios}) is consistent with models that predict that shock or X-ray processes heat the gas, rather than that the H$_{2}$ is thermally excited by UV photons (the latter would result in expected gas temperatures T\,$\la$1000\,K; see \citealt{mou94} and references therein). The 2-1\,S(1)/1-0\,S(1) ratios for both the narrow and broad component in regions A and B are clearly inconsistent with what is expected when non-thermal UV-fluorescence dominates the excitation process. Within the errors, the broad-component in both region A and B falls close to the temperature of T\,$\sim$\,2000 \,K along the LTE-line expected from uniform density gas that is fully thermalized. However, the deviation from this LTE-line indicates also here that the gas excitation may be more complex than simple LTE conditions. This is most prominent for the narrow component gas in region B, which suggests that either a non-negligible fraction of H$_{2}$ excitation through non-thermal processes or a mixing of molecular gas-components with different densities may occur for the H$_{2}$ gas in the large-scale disk. These results are consistent with the more detailed level population analysis of Fig.\,\ref{fig:tempfits} that we discussed above.

\subsubsection{Mass of the hot molecular gas}
\label{sec:mass}

In Sect. \ref{sec:temperature} we showed that the H$_{2}$ level populations of the broad-component emission in regions A and B are broadly consistent with hot molecular gas at a temperature of ${\rm T} \sim 1900 \pm 300$~K. This allows us to estimate the mass of the hot molecular gas following \citet{sco82} and \citet[][]{maz13}:
\begin{equation}
{\rm M}_{\rm hot} \simeq 5.0875 \times 10^{13} \left(\frac{D}{{\rm Mpc}}\right)^2 \left(\frac{I_{\rm 1-0S(1)}}{{\rm erg\,s^{-1}\,cm^{-2}}}\right) 10^{(0.4\,\cdot\,A_{2.2})}~{\rm M}_{\odot},
\label{eq:mass}
\end{equation}
where $D$ is the distance to the galaxy and $A_{\rm 2.2\mu m}$ the extinction at 2.2$\mu$m. This approach assumes thermalized gas conditions and T\,=\,2000\,K, with a population fraction in the ($\nu,J$) = (1,3) level of $f_{\rm (1,3)} = 0.0122$. As discussed in Sect.\,\ref{sec:temperature}, the assumption of LTE conditions for the broad-component H$_{2}$ gas is roughly consistent to within the accuracy of our level-population analysis, although uncertainties in this remain (for example, if a fraction of the H$_{2}$ gas would be sub-thermally excited, $f_{1,3}$ would be lower and equation \ref{eq:mass} would thus give lower limits to the H$_{2}$ mass).

According to \citet{piq13} A$_{2.2\mu \rm m} \approx 0.1$\,$\times$\,A$_{\rm V}$, with A$_{\rm V}$ the visual extinction. Because the broad-component H$_{2}$ emission is found outside the heavily obscured nuclei, we calculate  A$_{\rm V}$ from the Br$\gamma$/Br$\delta$ ratio in our spectra in regions A, B and C, following \citet{piq13}. For region B we derive A$_{\rm V} = 4.3 \pm 1.3$ mag, which is close to the median A$_{\rm V} = 5$ in NGC\,3256 found by \citet{piq13}. For region A the Br$\gamma$ and Br$\delta$ lines are too weak to derive reliable estimates, while for region C we derive A$_{\rm V} = 0.5 \pm 0.5$ mag. If we assume A$_{\rm V} = 4.3$ in both region A and B, then A$_{2.2\mu \rm m} \approx 0.1 \times$ A$_{\rm V} \sim 0.43$. Applying this A$_{2.2\mu \rm m}$ results in a mass estimate of the {\sl hot} molecular gas of M$_{\rm hot-H_2} \sim 630$ and 570 M$_{\odot}$ for the broad-component emission in region A and B, respectively. We note that when taking the strict lower limit of A$_{\rm V} = 0$ in region A, the corresponding mass estimate would lower by only $\sim 33\%$. 

Concluding, we estimate that the combined mass of hot molecular gas associated with the broad-component emission in regions A and B is M$_{\rm hot-H_2} \sim 1200$ M$_{\odot}$.

\subsection{Cold molecular counterpart}
\label{sec:ALMA}

   \begin{figure*}
   \centering
   \includegraphics[width=0.83\hsize]{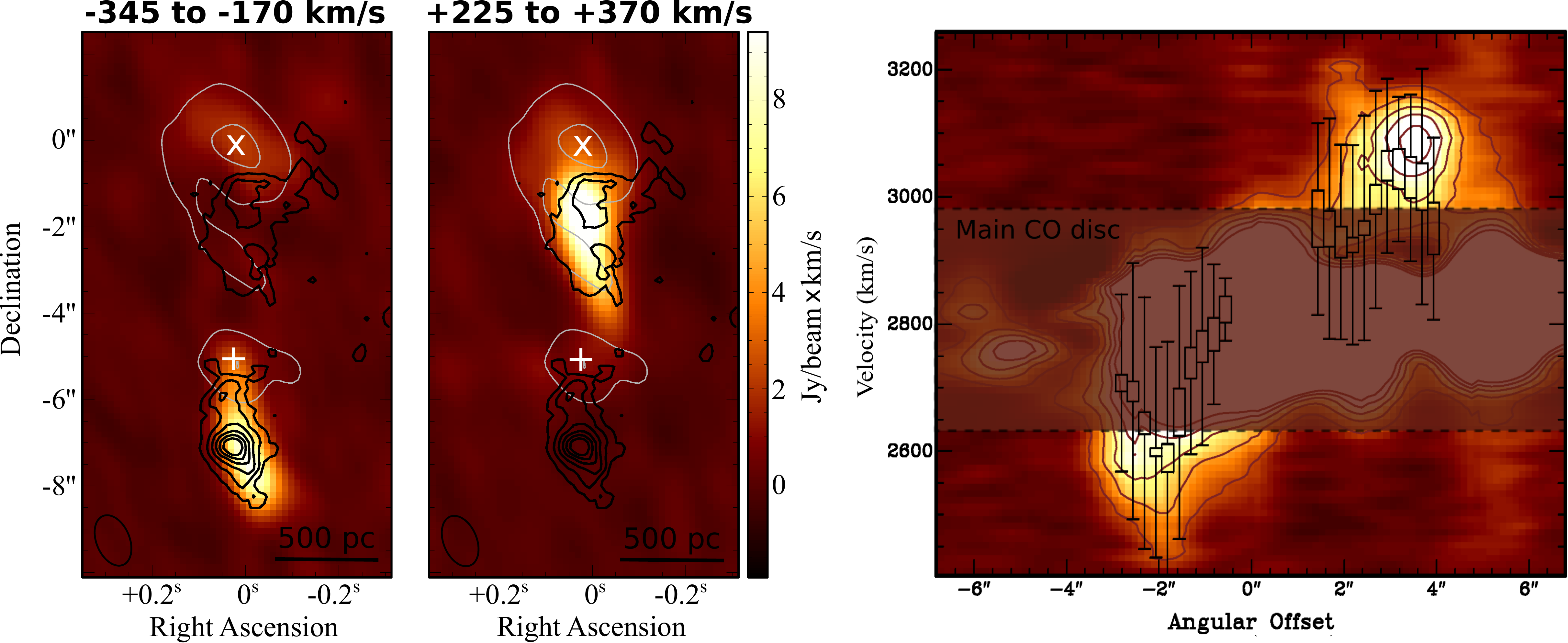}
      \caption{Total intensity map of the blueshifted {\sl (left)} and redshifted {\sl {(middle)}} high-velocity CO(3-2) emission in ALMA observations of NGC\,3256. The velocity range of the integrated CO(3-2) emission is given in the plots. Overlaid are thick black contours of the broad-component H$_{2}$ flux from Fig.\,\ref{fig:data} (levels from 14$\%$ to 80$\%$ in steps of 13$\%$ of the peak flux), as well as thin gray contours of the 343\,GHz radio continuum from the ALMA data \citep[levels at 3 and 10 mJy\,beam$^{-1}$; for details see][]{sak14}. The relative astrometry between the SINFONI and ALMA data has an estimated uncertainty of $\sim 0.5''$. {\sl Right:} Position-Velocity (PV) diagram of the high-velocity CO(3-2) emission along a north-south direction crossing the secondary nucleus at offset = 0$'$$'$ (optical velocity definition; contours 20, 40, 60, 80, 100 mJy\,beam$^{-1}$). Overlaid are black symbols that represent the broad-component H$_{2}$ emission along the same direction (i.e., the NS-aperture from Fig,\,\ref{fig:data}); the box shows the range between the average and maximum value per resolution element (1.1$''$ $\times$ 0.25$''$), while the error bars indicate $\pm$0.5$\times$FWHM$_{\rm average}$. The complex region of the main CO disk is not discussed (see \citealt{sak14}).}
      \label{fig:ALMA}
   \end{figure*}

In Fig.\,\ref{fig:ALMA} we compare our broad-component H$_{2}$ emission with high-velocity CO(3-2) gas (i.e., gas with velocities above those of the main CO disk), as observed with ALMA by \citet{sak14}. Fig.\,\ref{fig:ALMA} illustrates that the broad-component H$_{2}$ emission of {\sl hot} molecular gas in regions A and B has a counterpart in the {\sl cold} molecular gas as traced by CO(3-2). The H$_{2}$ and CO show a {\sl similar morphology and kinematics}, which are distinctly different from that of the main CO disk (which has its kinematic axis in east-west direction at position angle $\sim$75$^{\circ}$ and inclination $\sim$30$^{\circ}$; see \citealt{sak14} and Sect.\,\ref{sec:nature}). There could be a slight off-set between the location of the peak intensity for the H$_{2}$ and CO(3-2) emission in regions A and B, but this is within the uncertainty of our relative astrometry. No CO counterpart of the broad-component H$_{2}$ feature in region C was reliably distinguished from the main CO disk, although this is possibly due to the relatively small $\Delta$v$_{\rm peak}$ and $\sigma_{\rm max}$ that the broad-component molecular gas will have in this region (Table \ref{tab:results}).

We measure CO intensities of $I_{\rm CO(3-2)}$ = 23.5 and 39.7 Jy \kms\ in region A and B (Fig.\,\ref{fig:ALMA}). From this, we can derive mass estimates of the cold molecular gas if we know the excitation conditions of the gas. The high-velocity gas was also detected in CO(1-0) by \citet{sak14} at lower spatial resolution. Through a basic comparison of the high-velocity component in the CO(3-2) and CO(1-0) data, we estimate that $L'_{\rm CO(3-2)}$/$L'_{\rm CO(1-0)}$ is close to unity for regions A and B, and thus broadly consistent with thermal excitation conditions. This takes into account that for region B \citet{sak14} estimate that $\sim$34$\%$ of the high-velocity CO(3-2) emission and $\sim$72$\%$ of the CO(1-0) emission is contaminated by a CO outflow originating from the primary nucleus. If we assume $L'_{\rm CO(3-2)}$/$L'_{\rm CO(1-0)} = 1$ and we exclude the $\sim$34$\%$ contamination of high-velocity CO(3-2) emission from the primary nucleus in region B, we can set a lower-limit estimate to the luminosity of the cold molecular gas of $L'_{\rm CO} \sim 1.3$ K km\,s$^{-1}$ pc$^{2}$ for region A and $L'_{\rm CO} \sim 1.4$ K km\,s$^{-1}$ pc$^{2}$ for region B \citep[see][]{sol05}. By adopting a standard conversion factor for ULIRGs of $\alpha_{\rm CO} = {\rm M}_{\rm H2}/L'_{\rm CO} = 0.8$ \citep[][]{dow98}, this translates into a conservative estimate of the {\sl cold} molecular gas of M$_{\rm cold-H2} \approx 1.0$ and $1.1 \times 10^{7}$ M$_{\odot}$ for region A and B respectively, or M$_{\rm cold-H2} \approx 2.1$ $\times$ 10$^{7}$ M$_{\odot}$ in total. \citet{sak14} use the CO(1-0) intensity to estimate a total mass of 4.1 $\times$ 10$^{7}$ $X_{20}$ M$_{\odot}$ for this high-velocity gas (assuming $X_{20} \equiv \frac{N_{\rm H_2}/(1 \times 10^{20})}{I_{\rm CO(1-0)}}$ and adjusted for our assumed D=44.6 Mpc to NGC\,3256). Our assumption of $\alpha_{\rm CO} = 0.8$ translates to $X_{20} \sim 0.4$ \citep[see][]{bol13}. Therefore, taking into consideration the uncertainties involved regarding contamination of outflowing gas from the primary nucleus, our adopted $\alpha_{\rm CO} = 0.8$, and the gas thermalization, the two approaches agree to a mass of at least M$_{\rm cold-H2} \sim 2$ $\times$ 10$^{7}$ M$_{\odot}$ of cold molecular gas associated with the broad-component H$_{2}$ feature in regions A and B. We will use this value throughout the rest of this paper.

For further details on the CO data we refer to \citet{sak14}.

\section{Discussion}
\label{sec:discussion}

We presented broad-component H$_{2}$ emission-line features of hot molecular hydrogen gas in the vicinity of the heavily obscured secondary nucleus of NGC\,3256. We here discuss the nature of the broad-component H$_{2}$ emission in regions A and B. We present three possible scenarios: a biconical molecular gas outflow, a rotating disk that revolves around the secondary nucleus and a pair of tidal features. We will show that a molecular gas outflow best explains the observed properties (Sect.\,\ref{sec:nature}). The fact that this outflow is detected at high spatial resolution in both the hot and cold molecular gas phase makes it one of the few molecular outflows for which we can study in detail the physical structure and properties (Sect.\,\ref{sec:comparison}). We will also investigate the driving mechanism of the outflow (Sect.\,\ref{sec:mechanism}) and conclude with a brief remark about how our results fit into the broader perspective on molecular gas outflows (Sect.\,\ref{sec:lowres}).

\subsection{Nature of the broad-component H$_{2}$ emission}
\label{sec:nature}

\begin{figure}
   \centering
   \includegraphics[width=\hsize]{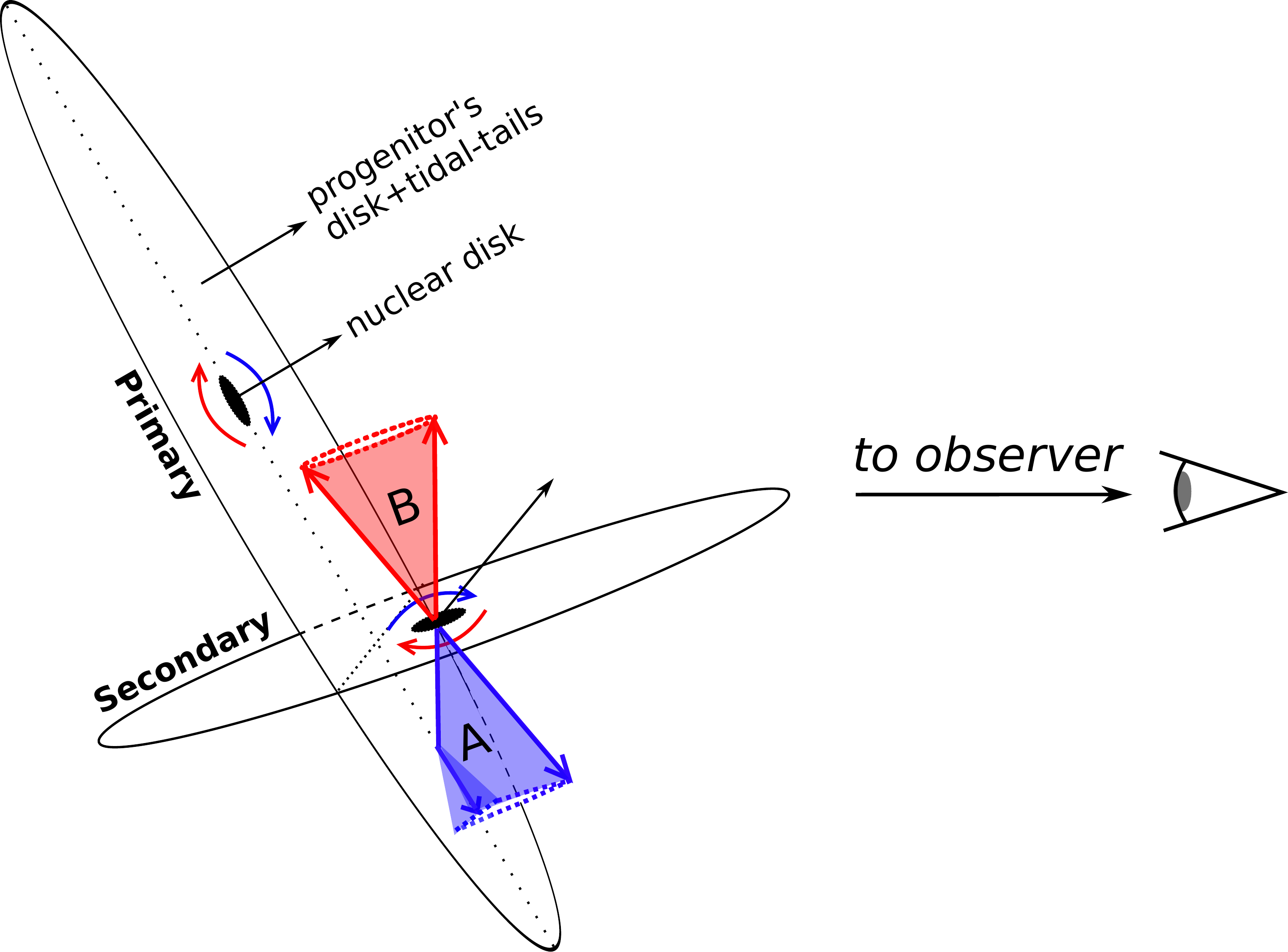}
      \caption{Alignment between the (nuclear) disks of the progenitor galaxies associated with the primary/secondary nucleus and the H$_{2}$ outflow. The alignment of the nuclear disks is based on CO observations by \citet{sak14}. Simplistically, the large-scale disks and tidal tails of the progenitor galaxies are assumed to be aligned parallel to the nuclear disks.}
      \label{fig:sketch}
   \end{figure}

The first scenario to explain the broad-component H$_{2}$ emission in regions A and B is that of a kpc-scale biconical outflow of molecular gas that originates from the secondary nucleus. We will here show that this scenario explains well the geometry and kinematics of the observed H$_{2}$ features. In Fig.\,\ref{fig:sketch} we visualize the geometry of the large-scale molecular gas disk and the biconical outflow from the secondary nucleus, as suggested by \citet{sak14}. The ALMA CO data of \citet{sak14} reveal that the large-scale disk has an inclination of $\sim$30$^{\circ}$ and that the southern part of the large-scale disk is the near side if the tidal-tails are trailing. They also show that the orbital plane of the primary and secondary nucleus must be close to this large-scale gas disk, but that the secondary is in front of the primary core and the secondary's nuclear disk has an inclination of at least $i_{\rm disk} \sim 70^{\circ}$. If we assume that the outflow is perpendicular to the nuclear disk, this would render the molecular outflow close to the plane of the sky ($i_{\rm outfl} \sim 20^{\circ}$). A biconical molecular gas outflow in a system with this geometry would explain why we see blueshifted H$_{2}$ emission in region A and redshifted emission in region B. In addition, the geometry in Fig.\,\ref{fig:sketch} suggests that the outflow may interact with the large-scale disk in region A. If this interaction results in shock-heating of molecular gas, this can naturally explain the fact that in region A the enhanced flux of H$_{2}$ emission in the outflowing broad-component gas is co-spatial with an enhancement in the H$_{2}$ flux of the narrow-component gas disk (Fig.\,\ref{fig:data}). Concluding, the observed geometry and kinematics of the broad-component H$_{2}$ emission agree with what is expected from a molecular outflow.

A second scenario is that the broad-component H$_{2}$ emission is an edge-on molecular disk or ring that rotates around the secondary nucleus. The velocity dispersion of the broad-component H$_{2}$ emission in NGC\,3256 is distinctively larger than that of the main gas disk (Sect.\,\ref{sec:results}), with v/$\sigma \sim 1.3$. This is intermediate between that of rotation-dominated (v/$\sigma$$\ga$1) and random-motion-dominated (v/$\sigma$$\la$1) systems \citep[e.g.,][]{epi12,bel13}, suggesting resemblance to a ``thick disk'', rather than a classical ``thin disk''. However, two arguments make this scenario unlikely. First, the dynamical mass enclosed by this rotating structure would be M$_{\rm dyn} \approx \frac{{\rm v_{\rm rot}^2}\,{\rm R}}{{\rm G}} \sim 1 \times 10^{10}$ M$_{\odot}$, although following \citet{bel13} it could be as high as M$_{\rm dyn} \approx \frac{2\,{\rm R_{\rm eff}}\,\left({\rm v_{\rm rot}^2}\,+\,1.35\,\sigma^{2}\right)}{{\rm G}} \sim 3 \times 10^{10}$ M$_{\odot}$ if we take into account the velocity dispersion of the gas (assuming v$_{\rm rot}$\,=\,$\Delta$v$_{\rm peak}$\,=\,250 \kms, $\sigma$\,=\,190 \kms, R\,=\,700\,pc and G\,=\,6.67\,$\times$10$^{-11}$ m$^{3}$\,kg$^{-1}$\,s$^{-2}$ the gravitational constant; Table\,\ref{tab:results}). This would result in a gas mass to dynamical mass ratio of M$_{\rm gas}$/M$_{\rm dyn} \approx 1/500 - 1/1500$. This is two orders of magnitude below the typical value of M$_{\rm gas}$/M$_{\rm dyn} \approx 1/6$ observed for the main central molecular gas disks of ULIRGs \citep[][]{dow98}. Second, a rotating disk or ring would cross the large-scale primary disk, which would trigger cloud-cloud collisions that would make the structure very unstable and short-lived. We thus argue that a rotating disk or ring is not a likely alternative explanation for the broad-component H$_{2}$ emission in regions A and B.

A third scenario is that the broad-component H$_{2}$ emission in regions A and B represents tidal features that are not in dynamical equilibrium. The complex kinematics observed within NGC\,3256 make this an attractive alternative scenario to investigate. The interaction between the putative tails and the ISM in regions A and B could increase the velocity dispersion of the gas within the tails. However, the observed broad-component features in regions A and B are only detected in the molecular gas phase and have no clear stellar or dusty counterparts, as would be expected for tidal tails (Fig.\,\ref{fig:NICMOS}). In addition, the spatial and kinematic geometry of the broad-component structure with respect to the secondary nucleus strongly suggests that the tidal gas would have been stripped off the secondary's central disk. However, \citet{sak14} show that the merger orbital plane is close to face-on and they therefore argue that the tidal force between the primary and secondary nucleus cannot produce large line-of-sight velocities for gas that is stripped off the secondary's central disk. Thus, the currently available data do not seem to favor the scenario that the broad-component H$_{2}$ emission reflects tidal features.

\subsection{Detailed physical properties of the outflow}
\label{sec:comparison}

In Sect.\,\ref{sec:nature} we argued that a biconical molecular gas outflow best explains the observed properties of the broad-component H$_{2}$ emission in regions A and B. The high spatial resolution of our data reveals interesting physical properties of the outflow. From the narrow width of the H$_{2}$ outflow in EW-direction ($\la 300$pc), our data indicate that the outflow has an opening angle of $\sim$40$^{\circ}$. The highest-velocity H$_{2}$ gas is found 440 (800) pc south (north) of the secondary nucleus. This suggests that the molecular outflow peaks far from the nucleus, possibly resembling a biconical shell-like geometry, with the lower-velocity molecular gas found down-stream of this shell.

The hot-to-cold molecular gas mass ratio of the outflowing gas is $\sim$\,$6 \times 10^{-5}$, which is at the high end of the range 10$^{-7}$\,-\,10$^{-5}$ observed across a sample of several dozen star-forming galaxies and AGN by \citet{dal05}. Moreover, H$_{2}$ has a short cooling timescale of $\sim$10$^{4}$ yr, which is two orders of magnitude shorter than the gas outflow timescale that we will calculate below (t$_{\rm outfl} \sim 10^{6}$ yr). This suggests that molecular gas in the outflow may be continuously shocked or heated (as is also observed to be the case for a molecular outflow from the radio-loud ULIRG PKS\,1345+12; \citealt{das14}.) The fact that the outflow appears to be predominantly molecular (Sect.\,\ref{sec:results}) suggests that the gas is most likely entrained in a wind from the AGN/starburst region, possibly by dragging molecular gas out of the secondary's circum-nuclear disk. This gas is likely heated through relatively slow shocks or X-rays (potentially from the AGN) to our observed temperature of $\sim$1900\,K. It is not immediately apparent that the H$_{2}$ emission could represent a transient phase of gas-cooling after fast shocks accelerate and heat/ionize gas directly at the shock-front (as appears to happen in the nearby active galaxy IC\,5063; \citealt{tad14}), because we do not see evidence of copious amounts of ionized gas in Br$\gamma$ that one would expect to be created by such fast shocks.

In Sect.\,\ref{sec:results} we found that the kinematic signature of the outflowing H$_{2}$ gas has an observed FWZI\,$\approx$\,1200 \kms, which means that the molecular gas in the outflow reaches a maximum line-of-sight velocity of $\pm$600 \kms. If we follow the outflow geometry described in Sect.\,\ref{sec:nature} and Fig.\,\ref{fig:sketch}, with $i_{\rm outfl} \sim 20^{\circ}$, this means that the intrinsic outflow velocities can reach a maximum of about $\pm$\,1800\,\kms. Similarly, if we assume our observed average outflow velocity to be v$_{\rm avg} = 0.5 \times {\rm FWHM_{\rm tot}} \approx 280$ \kms, this translates into an intrinsic average velocity of v$_{\rm avg} \sim 280$ {\rm sin}$^{-1}(i_{\rm outfl}) \sim 820$\,\kms. If we also de-project the outflow radius to $R = 700$ cos$^{-1}(i_{\rm outfl})$ $\sim 750$ pc, this means that it would take the gas roughly $10^{6}$ yr to travel this distance. We note that \citet{sak14} argue that the inclination of the outflow axis could even be as low as $i_{\rm outfl} \sim 5^{\circ}$, which would increase the above mentioned intrinsic outflow velocities with a factor of $\sim$4. \citet{sak14} base this low value for $i_{\rm outfl}$ on the fact that blue- and redshifted CO emission are observed to be co-spatial within the conical outflow, which they argue means that the inclination of the outflow is less than half the opening angle of $\sim$20$^{\circ}$ that they measure from their CO data (i.e., $i_{\rm outflow} \le 10^{\circ}$). However, this does not take into account the large velocity dispersion that we measure for the H$_{2}$ gas, which could explain the co-spatial red- and blueshifted emission even when $i_{\rm outfl} > 10^{\circ}$. In addition, we find an opening angle for the outflow of $\sim$40$^{\circ}$, larger than the $\sim$20$^{\circ}$ that \citet{sak14} derived from the CO data. We therefore argue that $i_{\rm outfl} < 20^{\circ}$ is not required to explain the observed properties of the molecular outflow.

With an intrinsic v$_{\rm avg} \sim 820$~\kms, $R \sim 750$\,pc and a total molecular gas mass of $\sim 2 \times 10^{7}$ M$_{\odot}$ involved in the outflow, the average mass outflow rate is \.{M}$_{\rm outfl} \sim 20$ M$_{\odot}$\,yr$^{-1}$. This is similar to that of other molecular outflows observed in low-$z$ starburst galaxies and AGN \citep[e.g.,][see also Sect.\,\ref{sec:intro}]{ala11,mor13,cic14,gar14}. This value for \.{M}$_{\rm outfl}$ is likely a lower limit, because our estimate of \.{M}$_{\rm outfl}$ is based on the over-simplified assumption that the total gas mass is driven out in a single event. As we discussed above, it is possible that the outflow is shell-like, or at least continuously re-filled with outflowing clouds, in which case \.{M}$_{\rm out}$ can be larger by at least a factor of a few \citep[see][]{mai12,cic14}. Still, \.{M}$_{\rm outfl} \sim 20$ M$_{\odot}$\,yr$^{-1}$ is high compared to the star-formation rate of the secondary nucleus. Following to \citet{kot96} and \citet{piq14}, the star-formation rate in the secondary nucleus is SFR $\sim 1-3$ M$_{\odot}$ yr$^{-1}$ (corrected for estimated extinctions of A$_{\rm V} \sim 10 - 12$ mag). Although significant uncertainty remains in these estimates, the mass-loading factor could be as high as $\eta$\,=\,\.{M}$_{\rm out}$/SFR\,$\sim$\,10.

\subsection{Outflow mechanism: AGN vs. starburst}
\label{sec:mechanism}

The relatively high mass loading factor of the outflow ($\eta$\,=\,\.{M}$_{\rm out}$/SFR\,$\sim$\,10) indicates that there is a discrepancy between the mass outflow rate and star-formation rate. A similar result on a molecular outflow in the nearby Seyfert galaxy NGC\,1068 led \citet{gar14} to favor AGN activity over star formation as the likely mechanism to drive the outflow (based on work by \citealt{mur05} and \citealt{vei05}). Moreover, the morphology of the H$_{2}$ outflow from the secondary nucleus in NGC\,3256 reveals that the outflow is rather collimated with an opening angle of $\sim$40$^{\circ}$, while the instrinsic maximum outflow velocity is high (v$_{\rm max}$\,$\sim$\,1800\,\kms). \citet{arr14} show that AGN in (U)LIRGs generate ionized gas outflows that are twice as fast as those produced by starbursts in these systems. These results could suggest that a hidden AGN may contribute to driving the outflow, for example through radiation pressure or a magnetically driven accretion-disk wind \citep[e.g.,][]{eve05,pro07}. \citet{sak14} also show marginal evidence for the presence of a faint two-sided radio-jet emanating from the secondary nucleus, which appears to be aligned with the high-velocity CO(3-2) emission \citep[based on data from][]{nef03}. 
Still, unambiguous observational evidence for the presence of an AGN in the secondary nucleus is currently lacking (Sect.\,\ref{sec:intro}). So are at least the energetics of the outflow consistent with the scenario that a hidden AGN, or else the nuclear starburst, can drive the outflow? 

The combined turbulent and bulk kinetic energy of the molecular outflow,
\begin{equation}
E_{\rm tot} = E^{\rm turb}_{\rm kin} + E^{\rm bulk}_{\rm kin} = {\frac{3}{2}} \cdot {\rm M}\sigma^{2} + {\frac{1}{2}} \cdot {\rm M}{\rm [v/sin({\it i}_{outfl})]}^{2},
\end{equation}
is $E_{\rm tot} \sim 2 \times 10^{56}$ erg (for M $\approx$ $2 \times 10^{7}$ M$_{\odot}$, $\sigma \approx 190$ \kms, ${\rm v} \approx 280$ \kms\ and $i_{\rm outfl} = 20^{\circ}$). We first compare this to the radiation from a potential AGN accretion disk. The $0.5-10$ keV X-ray luminosity from the secondary nucleus is $L_{\rm X} \sim 3 \times 10^{40}$ erg\,s$^{-1}$ \citep[assuming a power-law spectral model with $\Gamma = 2.0$ and an absorbing column of $5 \times 10^{22}$ cm$^{-2}$;][]{lir02}. \citet{elv94} show that QSOs with $L_{\rm 1\,-\,10\,keV} \sim 10^{43-47}$ erg\,s$^{-1}$ have a bolometric luminosity $L_{\rm bol}$ that is $5 - 50 \times$ larger than $L_{\rm 1\,-\,10\,keV}$, with an average/median value of $L_{\rm bol}/L_{\rm 1\,-\,10\,keV} \sim 20$. \citet{ho08} estimate a similar value of $L_{\rm bol}/L_{\rm 2\,-\,10\,keV} \sim 16$ for low-luminosity AGN. \citet{lir02} show that most of the $0.5-10$ keV X-ray emission from the secondary nucleus of NGC\,3256 occurs in the hard $2 - 10$ keV regime, so if we simplistically assume that the X-ray emission comes from a hidden AGN and we apply $L_{\rm bol}/L_{\rm X} \sim 16$, then the bolometric AGN luminosity would be $L_{\rm bol} \sim 5 \times 10^{41}$ erg\,s$^{-1}$ in case the AGN would be X-ray$-$transparent or Compton-thin. Over the outflow timescale of 10$^{6}$ yr, the total energy deposited by the AGN would be $\sim 2 \times 10^{55}$ erg\,s$^{-1}$, an order of magnitude too low to drive the outflow. However, if the X-ray emission is scattered emission from a Compton-thick (i.e., X-ray$-$obscured) AGN, the AGN's intrinsic $L_{\rm X}$, and thus also $L_{\rm bol}$, can be be a factor of $60 - 70$ higher \citep{pan06,sin11}. The non-detection of the Fe\,$\alpha$ line at 6.4 keV has led \citet{per11} to conclude that the upper limit to $L_{\rm bol}$ for a Compton-thick AGN is on the order of 10$^{43}$ erg\,s$^{-1}$ in NGC\,3256. Still, when adopting this upper limit, it would imply that a Compton-thick AGN, in case it has remained active over the past $\sim$10$^{6}$ yr, may have released a total energy of up to $\sim 3 \times 10^{56}$ erg\,s$^{-1}$ over this period, which would have been sufficient to drive the outflow.

How about mechanical energy from the putative radio source? As shown by \citet[][their Fig.\,19]{sak14}, the weak radio source appears to have a two-sided jet with a total 8\,GHz radio flux of at most $F_{\rm 8\,GHz} \sim 1$\,mJy (i.e., not taking into account the nuclear emission). From the flux of the jet we can derive the bulk kinetic power of the radio source by following \citet{wil99} and \citet{god13}:
\begin{equation}
Q_{\rm jet} \approx f^{3/2} ~ 3 \times 10^{38} \left( \frac{L_{\rm 151\,MHz}}{10^{28}\,{\rm W\,Hz^{-1}\,sr^{-2}}} \right) ~ {\rm W},
\end{equation}
where $L_{\rm 151\,MHz}$ is the 151 MHz luminosity (derived from the 151\,MHz flux and luminosity distance through $L_{\rm 151\,MHz} = F_{\rm 151}\,D_{\rm L}^{2}$) and $f$ is a factor that represents errors in the model assumption, including the excess energy in particles compared with that in the magnetic field, geometrical effects and energy in the backflow of the lobe (see \citealt{wil99} and \citealt{blu00} for details). Following \citet{blu00}, the value of $f$ is typically around $10-20$, so here we assume $f = 15$. In the optimistic case that the jet has a steep radio spectrum with $S_{\nu} \propto \nu^{-1}$, thus $F_{\rm 151} \sim 55$\,mJy, we estimate that Q$_{\rm jet} \sim 2 \times 10^{40}$\,erg\,s$^{-1}$. Over a typical radio-source lifetime of 10$^{6}$ yr, the mechanical energy deposited by the radio source would only be on the order of $\sim 6 \times 10^{53}$ ergs. This is two orders of magnitude lower than the combined turbulent and bulk kinetic energy of the molecular outflow, which suggests that the putative radio source currently does not have the power that is required to drive the outflow. Still, the presence of a radio source carving its way through the ISM may potentially have aided in creating a cavity through which molecular gas can be efficiently accelerated.

Alternatively, can the necessary energy for driving the outflow be injected by a nuclear starburst? \citet{kot96} and \citet{nor95} derive a supernova-rate in the secondary nucleus of $R_{\rm SN}$ $\sim$ 0.3 SN\,yr$^{-1}$. Assuming an energy-release per supernova event of $\sim$ 10$^{51}$ erg \citep{bet90}, and the fact that up to 90$\%$ of the energy per supernova is likely radiated away \citep{tho98}, the total energy deposited into the ISM by supernovae over the course of the outflow timescale $t_{\rm outfl}$ is roughly
\begin{equation}
E_{\rm SN} = 10^{51} \cdot \epsilon \cdot R_{\rm SN} \cdot t_{\rm outfl}~~{\rm (erg)},
\end{equation}
with $\epsilon$ the assumed efficiency of energy transfer to the ISM. For $R_{\rm SN}$ $\sim$ 0.3 SN\,yr$^{-1}$, $\epsilon \sim 0.1$ and $t_{\rm outfl} \sim 10^{6}$ yr, the energy-deposition by supernovae is $E_{\rm SN} = 3 \times 10^{55}$ ergs. Stellar winds will release additional energy into the system, although \citet{lei92} show that once the starburst has aged enough that supernova-explosions occur, stellar winds will increase the total energy deposited into the ISM by at most a factor of 2, putting a limit of $E_{\rm SN+winds} \sim 6 \times 10^{55}$ ergs. This is only a factor of 3 lower than the energy of the outflow. Given the assumptions involved, and the fact that it is uncertain to what extent the large amounts of radiative energy may contribute to driving the outflow, it is possible that the energy output of the starburst event in the secondary nucleus may be sufficient to sustain the outflow.

Concluding, the high mass loading factor, high outflow velocities and substantial collimation of the outflow suggest that that a hidden AGN may contribute to driving the outflow of molecular gas from the secondary nucleus. Based on the energetics, we argue that the most plausible way a hidden AGN can drive the outflow is in the Compton-thick regime, through radiation-pressure or an accretion-disk wind. However, the scenario that the nuclear starburst provides the energy needed to drive the outflow is also possible.

\subsection{Broader perspective}
\label{sec:lowres}
 
Interestingly, while the biconical structure of the outflow shows a distinct blue- and redshifted wing to the emission-line profile on either side of the secondary nucleus, when integrated across the entire outflow region the broad-component profile appears symmetric with respect to the assumed systemic velocity of the secondary nucleus (Fig.\,\ref{fig:datatotal}). This is similar to what is often seen in CO data with much lower spatial resolution, and for objects at intermediate- and high-$z$ \citep[e.g.,][]{fer10,mai12,cic14}. Thus, our results provide valuable insight into understanding the structural properties of starburst/AGN-driven molecular gas outflows in general.

\section{Conclusions}

We presented evidence for the presence of a kpc-scale biconical outflow of hot molecular gas from the heavily obscured secondary nucleus of NGC 3256, based on near-IR 2.12$\mu$m H$_{2}$ data obtained with VLT/SINFONI. Our main conclusions are:

\vspace{1.5mm}

\noindent {\sl i).} The outflow is observed at high spatial resolution in both the hot molecular gas phase with VLT/SINFONI (H$_{2}$) and the cold molecular phase with ALMA (CO). The hot and cold component of the molecular outflow share a similar morphology and kinematics. 

\vspace{1.5mm}

\noindent {\sl ii).} These data allowed us to characterize the geometry, kinematics and physical properties of the molecular outflow. In particular:
\begin{itemize}
\item{The outflow consists of a blueshifted outflow-component south and a redshifted component north of the secondary nucleus, with an opening angle of $\sim40^{\circ}$;}
\vspace{1.5mm}
\item{The emission-line kinematics show observed maximum outflow velocities of $\pm$600 \kms. However, give the low inclination of the jet-axis ($i_{\rm outfl} \sim 20^{\circ}$), intrinsic outflow velocities can reach a maximum of $\sim$\,1800\,\kms;}
\vspace{1.5mm}
\item{The mass of the hot molecular gas in the outflow is M$_{\rm hot\,H_2} \sim 1200$~M$_{\odot}$, while the outflowing cold molecular gas mass is M$_{\rm cold\,H_2} \sim 2 \times 10^{7}$\,M$_{\odot}$. This results in a hot-to-cold molecular gas mass ratio of $\sim$$6 \times 10^{-5}$ and total molecular mass-outflow rate of at least \.{M}$_{\rm outfl} \sim 20$ M$_{\odot}$\,yr$^{-1}$};
\vspace{1.5mm}
\item{From the analysis of multiple near-IR H$_{2}$-transitions, we derive a temperature of T\,$\sim$\,$1900\pm300$\,K for the hot H$_{2}$ gas in the outflow. The likely heating mechanism is either shocks or X-ray emission.} 
\end{itemize}

\vspace{0.8mm}

\noindent {\sl iii).} A likely driving mechanism for the molecular outflow is a hidden AGN, given the high mass-loading factor ($\eta \sim 10$), high outflow velocities and significant collimation of the outflowing gas. Based on energy requirements, this would most likely happen in the Compton-thick regime, through radiation pressure or an accretion-disk wind. Alternatively, the nuclear starburst may potentially provide enough energy to drive the outflow;

\vspace{1.5mm}

\noindent {\sl iv).} When integrated over the outflow region, the global kinematics of the outflowing molecular gas mimic those observed with low-resolution CO observations in other low- and high-$z$ objects. The structural and physical properties of the molecular outflow in NGC\,3256 that we derive from our high-resolution data therefore provide valuable insight into our general understanding of starburst/AGN-driven molecular gas outflows.

\begin{acknowledgements}
We thank the anonymous referee for valuable suggestions that substantially improved the content of this paper. BE is grateful that the research leading to these results has received funding from the Spanish Ministerio de Econom\'{i}a y Competitividad (MINECO) under grant AYA2010-21161-C02-01 and from the European Union Seventh Framework Programme (FP7-PEOPLE-2013-IEF) under grant agreement n$^{\circ}$\,624351. LC, SA and AAH acknowledge support through MINECO grants AYA-2012-39408-C02-01 and AYA-2012-31447. MPS is supported by the Agenzia Spaziale Italiana (ASI), contract I/005/11/0. Based on observations collected at the European Organisation for Astronomical Research in the Southern Hemisphere, Chile, prog. 078.B-0066A. The Atacama Large Millimeter Array (ALMA), an international astronomy facility, is a partnership of Europe, North America and East Asia in cooperation with the Republic of Chile. Based on observations made with the NASA/ESA Hubble Space Telescope, and obtained from the Hubble Legacy Archive, which is a collaboration between the Space Telescope Science Institute (STScI/NASA), the Space Telescope European Coordinating Facility (ST-ECF/ESA) and the Canadian Astronomy Data Centre (CADC/NRC/CSA).
\end{acknowledgements}


\appendix

\section{Properties H$_{2}$ transitions}
\label{app:H2lines}

In this Appendix we provide the spectra and line fluxes derived from fitting two Gaussian components to the emission-line spectra of the various H$_{2}$ transitions in regions A, B and C.

\subsection{Method}
\label{app:method}

For each of the seven H$_{2}$ transition in our SINFONI data we extracted an emission-line spectrum from a circular aperture of 13 spaxels centered around the peak flux in regions A, B and C (similar to what is shown in Fig.\,\ref{fig:data} for H$_{2}$ 1-0\,S(1)). 

As mentioned in Sect.\,\ref{sec:data}, we used an IDL routine based on the MPFIT package to perform the line-fitting. We used a three-component model to simultaneously fit the continuum and both the narrow and broad component of the emission lines. The continuum was fitted using a linear term, whereas a Gaussian profile was used for each emission-line component.

As can be seen in Fig.\,\ref{fig:app}, the signal-to-noise of some of the lines is too low to derive accurate results when using this multi-component analysis without placing any constraints on the line-fitting procedure. To reduce the uncertainties of the fitting and obtain more robust measurements of the line fluxes, we assume that all transitions share the same kinematics. We first performed the fitting routine on the strong 1-0\,S(1) 2.1218\,$\mu$m line without placing constraints on the line-fitting parameters. Subsequently, we used these results to fix the position and line width (both in \kms) of the narrow and broad component of the other emission lines. 

The uncertainties of the flux measurements were calculated using a Monte Carlo technique. This method consists of measuring the noise in the spectra as the root-mean-square of the residuals after the subtraction of our multi-component model. Taking this estimation of the noise into account, we construct a total of 500 independent simulations/realizations of the spectra, where the lines are again fitted. These simulations yield distributions of each free parameter of our model. The uncertainty of each parameter is then defined as the standard deviation of its corresponding distribution.

\subsection{Results}
\label{app:results}

\begin{figure*}
   \centering
   \includegraphics[width=0.8\hsize]{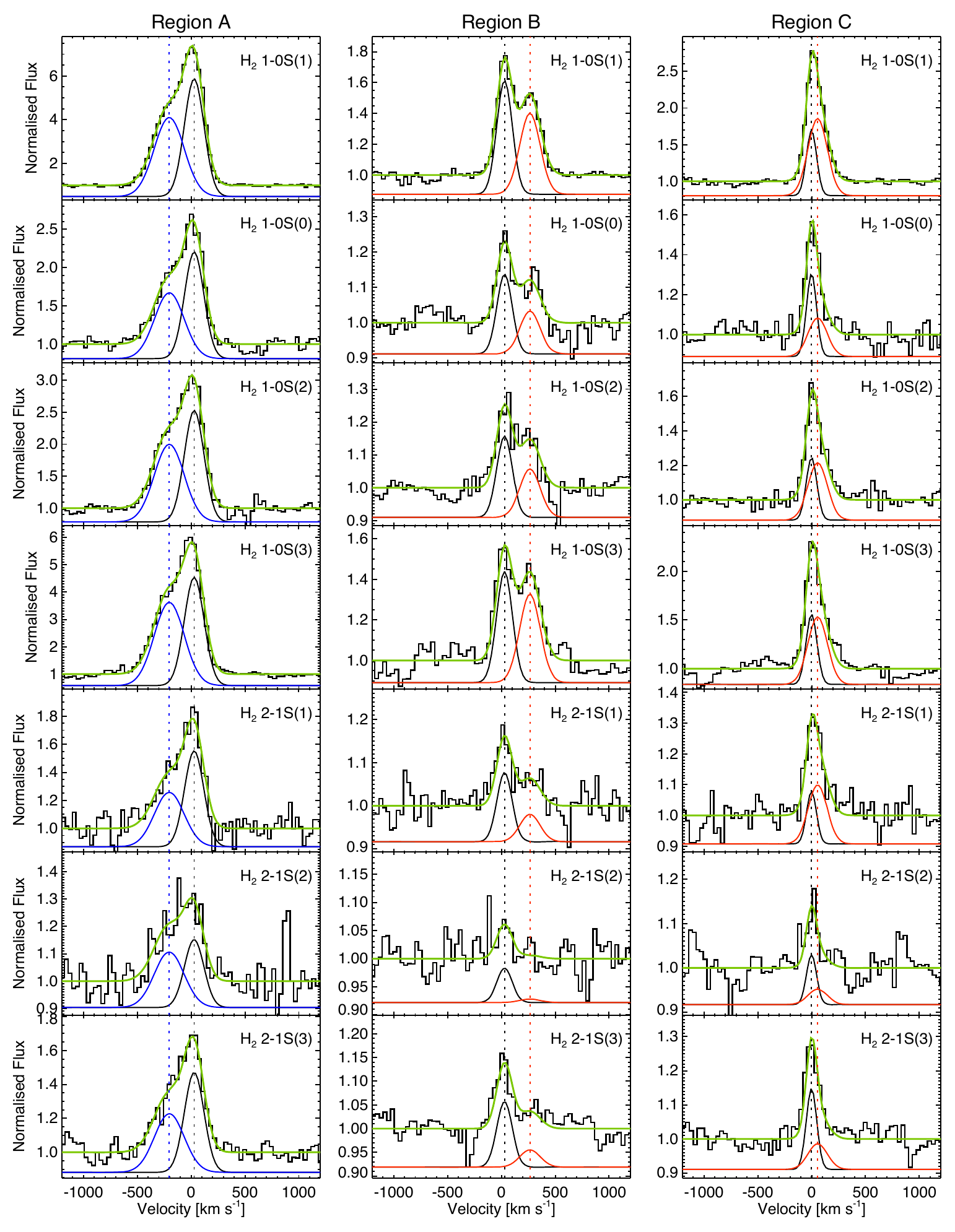}
      \caption{Emission-line spectra and 2-component Gaussian fit of the various H$_{2}$ transitions in regions A, B and C. The spectra were extracted from a circular aperture of 13 spaxels, centered on the broad-component H$_{2}$ feature in each region. The line width and shift between the peak of the narrow and broad component were constrained in velocity to those of the strong 1-0\,S(1) line. A straight line was fitted to the continuum across a wide velocity range of -1800 $-$ +1800 \kms\ in order to handle potential features in the continuum (see, e.g., the H$_{2}$\,1-0S(3) transition in region B). However, a smaller velocity range is plotted in the figures to high-light the details of the fits to the emission lines. After extracting flux measurements, for visualisation purposes the spectra in the plots were normalized to the fitted continuum.}
      \label{fig:app}
   \end{figure*}

In Fig.\,\ref{fig:app} we show the results of the line-fitting procedure. In Table \ref{tab:app} we summarize the line ratios for the broad and narrow component with respect to the flux of the 1-0\,S(1) 2.1218\,$\mu$m line. The upper limits are defined as 1-sigma detections, using the noise estimation from the Monte Carlo method described in Sect.\,\ref{app:method}. 

\begin{table*}
\caption{Flux ratios of the broad and narrow components of all seven H$_{2}$ lines in our SINFONI data with respect to 1-0\,S(1) in regions A, B and C. The absolute flux of the 1-0\,S(1) transition is provided at the top.}
\label{tab:app}
\begin{tabular}{cc|ccc|ccc}
H$_{2}$ trans. & $\lambda_{\rm rest}$\,($\mu$m) &  \multicolumn{3}{c|}{{\sl Broad component}} & \multicolumn{3}{c}{{\sl Narrow component}} \\
\hline
 & & & & & & \\
 & & \multicolumn{3}{c|}{Flux 1-0\,S(1) ($\times 10^{-16}$\,erg\,s$^{-1}$\,cm$^{-2}$)} & \multicolumn{3}{c}{Flux 1-0\,S(1) ($\times 10^{-16}$\,erg\,s$^{-1}$\,cm$^{-2}$)} \\
 & & A & B & C & A & B  & C \\
 1-0\,S(1) & 2.1218  & 5.966\,$\pm$\,0.306 & 1.913\,$\pm$\,0.188 & 2.135\,$\pm$\,0.399 & 5.826\,$\pm$\,0.297 & 2.028\,$\pm$\,0.175 & 0.953\,$\pm$\,0.403 \\
 & & & & & & & \\
 & & \multicolumn{3}{c|}{Flux ratio with respect to 1-0\,S(1)} & \multicolumn{3}{c}{Flux ratio with respect to 1-0\,S(1)} \\
 & & A & B & C & A & B  & C \\
 1-0\,S(0) & 2.2235  & 0.239\,$\pm$\,0.014 & 0.219\,$\pm$\,0.033 & 0.179\,$\pm$\,0.046 & 0.257\,$\pm$\,0.014 & 0.290\,$\pm$\,0.032 & 0.453\,$\pm$\,0.197 \\
 1-0\,S(1) & 2.1218  & 1.000\,$\pm$\,\,\,n.a.\,\,\, & 1.000\,$\pm$\,\,\,n.a.\,\,\, & 1.000\,$\pm$\,\,\,n.a.\,\,\, & 1.000\,$\pm$\,\,\,n.a.\,\,\, & 1.000\,$\pm$\,\,\,n.a.\,\,\, & 1.000\,$\pm$\,\,\,n.a.\,\,\, \\
 1-0\,S(2) & 2.0338  & 0.361\,$\pm$\,0.022 & 0.290\,$\pm$\,0.050 & 0.326\,$\pm$\,0.068 & 0.349\,$\pm$\,0.021 & 0.351\,$\pm$\,0.045 & 0.434\,$\pm$\,0.190 \\
 1-0\,S(3) & 1.9576  & 0.958\,$\pm$\,0.052 & 0.887\,$\pm$\,0.103 & 0.693\,$\pm$\,0.147 & 0.841\,$\pm$\,0.045 & 0.806\,$\pm$\,0.082 & 0.898\,$\pm$\,0.396 \\        
 2-1\,S(1) & 2.2477  & 0.109\,$\pm$\,0.010 & 0.113\,$\pm$\,0.027 & 0.175\,$\pm$\,0.043 & 0.128\,$\pm$\,0.010 & 0.203\,$\pm$\,0.026 & 0.181\,$\pm$\,0.089 \\
 2-1\,S(2) & 2.1542  & 0.054\,$\pm$\,0.008 & \multicolumn{1}{c}{$<$0.067} & \multicolumn{1}{c}{$<$0.049} & 0.045\,$\pm$\,0.006 & 0.080\,$\pm$\,0.021 & 0.123\,$\pm$\,0.065 \\
 2-1\,S(3) & 2.0735  & 0.101\,$\pm$\,0.008 & 0.070\,$\pm$\,0.021 & 0.072\,$\pm$\,0.025 & 0.115\,$\pm$\,0.007 & 0.190\,$\pm$\,0.023 & 0.275\,$\pm$\,0.121 \\
\end{tabular} 
\end{table*}








\end{document}